\documentclass{aastex62}

\graphicspath{{./}{figures/}}
\usepackage{color}

\shorttitle{Star density profile of six old LMC GCs}
\shortauthors{Lanzoni et al.}

\begin{document}

\title{Star density profiles of six old star clusters in the Large Magellanic Cloud}

\correspondingauthor{Barbara Lanzoni}
\email{barbara.lanzoni3@unibo.it}

\author[0000-0001-5613-4938]{Barbara Lanzoni}
\affil{Dipartimento di Fisica e Astronomia, Universit\`a di Bologna, Via Gobetti 93/2, Bologna I-40129, Italy}
\affil{Istituto Nazionale di Astrofisica (INAF), Osservatorio di Astrofisica e Scienza dello Spazio di Bologna, Via Gobetti 93/3, Bologna I-40129, Italy}

\author[0000-0002-2165-8528]{Francesco R. Ferraro}
\affil{Dipartimento di Fisica e Astronomia, Universit\`a di Bologna, Via Gobetti 93/2, Bologna I-40129, Italy}
\affil{Istituto Nazionale di Astrofisica (INAF), Osservatorio di Astrofisica e Scienza dello Spazio di Bologna, Via Gobetti 93/3, Bologna I-40129, Italy}

\author[0000-0003-4237-4601]{Emanuele Dalessandro}
\affil{Istituto Nazionale di Astrofisica (INAF), Osservatorio di Astrofisica e Scienza dello Spazio di Bologna, Via Gobetti 93/3, Bologna I-40129, Italy}

\author[0000-0002-5038-3914]{Mario Cadelano}
\affil{Dipartimento di Fisica e Astronomia, Universit\`a di Bologna, Via Gobetti 93/2, Bologna I-40129, Italy}
\affil{Istituto Nazionale di Astrofisica (INAF), Osservatorio di Astrofisica e Scienza dello Spazio di Bologna, Via Gobetti 93/3, Bologna I-40129, Italy}

\author[0000-0002-7104-2107]{Cristina Pallanca}
\affil{Dipartimento di Fisica e Astronomia, Universit\`a di Bologna, Via Gobetti 93/2, Bologna I-40129, Italy}
\affil{Istituto Nazionale di Astrofisica (INAF), Osservatorio di Astrofisica e Scienza dello Spazio di Bologna, Via Gobetti 93/3, Bologna I-40129, Italy}

\author[0000-0003-4592-1236]{Silvia Raso}
\affil{Dipartimento di Fisica e Astronomia, Universit\`a di Bologna, Via Gobetti 93/2, Bologna I-40129, Italy}
\affil{Istituto Nazionale di Astrofisica (INAF), Osservatorio di Astrofisica e Scienza dello Spazio di Bologna, Via Gobetti 93/3, Bologna I-40129, Italy}

\author[0000-0001-9158-8580]{Alessio Mucciarelli}
\affil{Dipartimento di Fisica e Astronomia, Universit\`a di Bologna, Via Gobetti 93/2, Bologna I-40129, Italy}
\affil{Istituto Nazionale di Astrofisica (INAF), Osservatorio di Astrofisica e Scienza dello Spazio di Bologna, Via Gobetti 93/3, Bologna I-40129, Italy}

\author[0000-0002-3865-9906]{Giacomo Beccari}
\affil{European Southern Observatory, Karl-Schwarzschild-Strasse 2, 85748 Garching bei Munchen, Germany}

\author{Paola Focardi}
\affil{Dipartimento di Fisica e Astronomia, Universit\`a di Bologna, Via Gobetti 93/2, Bologna I-40129, Italy}

\begin{abstract}
We used resolved star counts from {\emph Hubble Space Telescope}
images to determine the center of gravity and the projected density
profiles of 6 old globular clusters in the Large Magellanic Cloud
(LMC), namely NGC 1466, NGC 1841, NGC 1898, NGC 2210, NGC 2257 and
Hodge 11.  For each system the LMC field contribution was properly
taken into account by making use, when needed, of parallel {\emph HST}
observations. The derived values of the center of gravity may differ
by several arcseconds (corresponding to more dal 1 pc at the distance
of the LMC) from previous determinations.  The cluster density
profiles are all well fit by King models, with structural parameters
that may differ from the literature ones by even factors of two.
Similarly to what observed for Galactic globular clusters, the ratio
between the effective and the core radius has been found to
anti-correlate with the cluster dynamical age.
\end{abstract}


\date{5 November 2019}

\section{Introduction}
\label{sec_intro}
Globular clusters (GCs) are the best example in nature of collisional
stellar systems, where multiple and multifold gravitational
interactions occur among the constituent stars, significantly altering
the physical properties of the host with respect to its conditions at
formation \citep[see, e.g.,][]{meylan+97}.  In this context, we are
carrying on a long-term project aimed at the accurate characterization
of the internal structure, kinematics and stellar content of GCs.  In
particular, we are using number counts \citep{lanzoni+07a,
  lanzoni+07b, lanzoni+07c, lanzoni+10, miocchi+13}, in place of the
surface brightness distributions, and the radial velocities of
individual stars (\citealp{lanzoni+13, lanzoni+18a, lanzoni+18b,
  ferraro+18a}; see also \citealp{baum_hilker18}), instead of
  integrated-light spectroscopy, to determine the cluster
  gravitational centers and the structural and kinematical
  parameters. This is to avoid the so-called ``shot-noise bias'' that
  is known to affect luminosity-weighted quantities (as the surface
  brightness distribution and integrated-light spectra) when dealing
  with resolved stellar populations.  The bias is due to the
  stochastic and sparse presence of luminous stars, which can
  significantly displace the surface brightness peak from the true
  location of the cluster gravitational center, and alter the shape of
  the surface brightness profile with respect to the true density
  distribution (see, e.g., \citealt{noyola+06} for a discussion of
  methods adopted to correct for this problem in photometric studies,
  and see \citealt{dubath+97}, \citealt{lutzgendorf+11}, and
  \citealt{lanzoni+13} for a discussion of this bias in the case of
  integrated-light spectroscopy).  The bias does not occur if resolved
  stars are used, since every object has the same weight,
  independently of its luminosity \citep[e.g.][]{calzetti+93,
    lugger+95, montegriffo+95}.  In spite of their advantages,
  techniques based on star counts have not been fully exploited in the
  literature yet, and the vast majority of GC structural parameters
  listed in largely used catalogs (e.g., \citealp{h96};
  \citealp{mackey03}, hereafter MG03; \citealp{mclaugh05}, hereafter
  Mv05) have been derived from surface brightness distributions.  This
  is essentially because constructing complete samples of resolved
  stars in the highly crowded central regions of GCs is not an easy
  task. In fact, even {\emph Hubble Space Telescope (HST)} catalogs
  obtained from observations not optimized to avoid strong saturation
  from the bright giants can be severely incomplete in the central
  regions of high-density clusters \citep[see][]{ferraro+97, raso+17}.
  However, starting from studies dedicated to specific objects or very
  small sets of clusters \citep[e.g.][]{ferraro+99, ferraro+03, ema08,
    ema13, ema15, salinas+12, saracino+15, cadelano+17}, the
  systematic determination of gravitational centers and surface
  density profiles from resolved stars counts is increasingly adopted
  (see, e.g., the study of a sample of 26 Galactic GCs discussed in
  \citealt{miocchi+13}).

For a full physical characterization of these systems we are also
making use of the observational properties of special classes of
stellar objects, like the so-called ``blue straggler stars'' (BSSs;
e.g., \citealp{ferraro+06, ferraro+09, ferraro+12, ema13,
  beccari+19}). In fact, BSSs are significantly more massive than
normal cluster stars (e.g., \citealt{shara97, fiorentino14, raso+19})
and dynamical friction thus makes them progressively sinking toward
the cluster center. Correspondingly, their central segregation
relative to a reference (lighter) population (as horizontal branch,
red giant branch, main sequence stars) progressively increases with
time. This can be quantitatively measured from the shape of the BSS
radial distribution \citep[e.g.][]{ferraro+12}, and through the $A^+$
parameter, which is defined as the area enclosed between the
cumulative radial distribution of BSSs and that of the reference
population (\citealp{alessandrini16}; see also \citealp{lanzoni+16}
for a comparison between the two approaches). Recently,
\citet{ferraro+18b} measured the value of $A^+$ within one half-mass
radius from the center of 48 Galactic GCs ($\sim 32\%$ of the entire
Milky Way population) and found a strong correlation with the number
of central relaxation times ($N_{\rm rel}$) suffered by each system
since formation.  This demonstrates that $A^+$ is a powerful empirical
``dynamical clock'', able to efficiently measure the level of internal
dynamical evolution suffered by stellar systems (i.e., their dynamical
age).

We are now extending the same approach adopted in the Milky Way, to
star clusters located in the Large Magellanic Cloud (LMC).  This
nearby galaxy hosts stellar systems covering a wide range of ages
(from a few million, to several billion years), at odds with the Milky
Way where mostly old ($t>10$ Gyr) GCs are found. It therefore offers a
unique opportunity to explore the formation process of star clusters
over cosmic time, making the characterization of their physical
properties crucial. In addition, since the LMC tidal field differs
from that of the Milky Way, the dynamical evolution of the hosted
clusters could be different as well (see, e.g., \citealp{piatti19} for
a recent study of the effect of the Milky Way tidal field on GC
properties). \citet{ferraro+19} measured the $A^+$ parameter in five
old LMC clusters, finding that they follow the same correlations with
$N_{\rm rel}$ drawn by the Milky Way systems. Here we focus on the
determination of the projected density profiles and structural
parameters (from resolved star counts) of the same systems studied by
\citet{ferraro+19}, plus an additional one (NGC 1898) where the strong
contamination from LMC field stars prevented us from a safe selection
of the BSS population and the determination of $A^+$.

The paper is organized as follows. In Section \ref{sec_data} we
present the photometric database used and the adopted data reduction
procedures. In Sections \ref{sec_center} and \ref{sec_dens} we discuss
the determination of the cluster gravitational center and projected
density profiles from the observed resolved stars. Section
\ref{sec_models} is devoted to present the fit to the observed density
profiles through \citet{king66} models, and the derivation of the
cluster structural parameters. In Section \ref{sec_discuss} we
discuss the obtained results.

\begin{deluxetable}{l|l|l|l|c}
\tablecaption{Data set} \tablehead{\colhead{Cluster} &
  \colhead{Camera} & \colhead{Filter} & \colhead{Exposure Time} &
  \colhead{Prop. ID}} \startdata NGC 1466 & ACS/WFC & F606W & $2
\times 50$ s, $12 \times 353$ s & 14164\\ & ACS/WFC & F814W & $2
\times 70$ s, $6 \times 352$ s, $6 \times 385$ s, $6 \times 420$ s \\
parallel  & ACS/WFC   & F435W & $12\times 575$ s\\
          & ACS/WFC   & F606W & $1\times 50$ s, $3\times 566$ s\\
\hline
NGC 1841 & ACS/WFC   & F606W & $2 \times 50$ s, $12 \times 353$ s & 14164\\
         & ACS/WFC   & F814W & $2 \times 70$ s, $6 \times 352$ s, $6 \times 385$ s, $6 \times 420$ s \\
parallel  & ACS/WFC   & F435W & $12\times 575$ s\\
          & ACS/WFC   & F606W & $1\times 50$ s, $3\times 566$ s\\
\hline
NGC 1898 & ACS/WFC   & F475W & $2 \times 500$ s & 12257 \\
         & ACS/WFC   & F814W & $2 \times 500$ s \\
         & WFC3/UVIS & F336W & $2\times1035$ s  & 13435\\
\hline
NGC 2210  & ACS/WFC   & F606W & $2 \times 50$ s, $6 \times 348$ s, $6 \times 353$ s & 14164 \\
          & ACS/WFC   & F814W & $2\times70$ s, $6\times344$ s, $6\times378$ s, $6\times413$ s \\
          & WFC3/UVIS & F336W & $4\times700$ s, $4\times715$ s, $4\times729$ s, $4\times730$ s \\
parallel  & ACS/WFC   & F435W & $4\times 550$ s, $8\times 565$ s\\
          & ACS/WFC   & F606W & $1\times 50$ s, $3\times 560$ s\\
\hline
NGC 2257  & ACS/WFC   & F606W & $2 \times 50$ s, $3\times364$ s, $2\times525$ s, $6\times353$ s & 14164\\
          & ACS/WFC   & F814W & $2\times70$ s, $3\times390$ s, $2\times450$ s, $6\times363$ s, $6\times400$ s \\
parallel  & ACS/WFC   & F435W & $12\times 575$ s\\
          & ACS/WFC   & F606W & $1\times 50$ s, $3\times 570$ s\\
         \hline
Hodge 11  & ACS/WFC   & F606W & $2 \times 50$ s, $6\times345$ s, $6\times370$ s & 14164\\
          & ACS/WFC   & F814W & $2\times70$ s, $6\times345$ s, $6\times377$ s, $6\times410$ s \\
          & WFC3/UVIS & F336W & $3\times700$ s , $12\times729$ s \\
parallel  & ACS/WFC   & F435W & $4\times 550$ s, $8\times 570$ s\\
          & ACS/WFC   & F606W & $1\times 50$ s, $3\times 570$ s\\
         \hline
\enddata
\tablecomments{Details of the {\emph HST} archive images used in the present
  study.}
\label{tab_data}
\end{deluxetable}

\section{Photometric database and data reduction}
\label{sec_data}
The data used in this paper consist in a set of archive images
acquired with the Wide Field Channel of the Advanced Camera for Survey
(ACS/WFC) and the UVIS channel of the Wide Field Camera 3 (WFC3/UVIS)
on board the {\emph HST} (see Table \ref{tab_data}).  For five of the
selected clusters, the data are part of GO 14164 (PI Sarajedini) and
consist in ACS/WFC observations in the F606W ($V$) and F814W ($I$)
filters performed in the direction of each system, complemented with
parallel ACS/WFC images of nearby fields ($\sim 5\arcmin$ from the
cluster centers) acquired through the F435W ($B_{435}$) and F606W
filters.  In the case of NGC 2210 and Hodge 11 we also made use of the
WFC3/UVIS pointings in the F336W, acquired under the same program.
For NGC 1898 the images have been obtained with the ACS/WFC in the
F475W ($B$) and F814W filters (GO 12257, PI Girardi), and with the
WFC3/UVIS in the F336W (GO 13435, PI Monelli).  In general, different
pointings dithered by several pixels have been performed in each band,
thus allowing the filling of the inter-chip gaps, and an adequate
subtraction of CCD defects, artifacts, and false detections.

The photometric analysis was performed independently on each image via
the point-spread function (PSF) fitting method, by using
\texttt{DAOPHOT IV} and following the ``standard'' approach already
used in previous works (e.g., \citealp{ema18}).  Briefly, PSF models
were derived for each image and detector by using some dozens of
bright and isolated stars, and then applied to all the detected
sources with flux peaks at least $3\sigma$ above the local background.
A master list including stars detected in at least two images was then
created. At the position of all the stars in the master-list, a fit
was forced in each frame with \texttt{DAOPHOT/ALLFRAME}
\citep{stetson94}.  In the case of NGC 1898, NGC 2210 and Hodge 11,
the master list of detected sources has been constructed on the F336W
images, and then forced to the exposures acquired in the other
filters.  As quantitatively demonstrated in \citet{raso+17},
photometric analyses guided by short-wavelength filters allow the
optimization of the source detection in the overcrowded central
regions of old stellar populations, where cool giant stars easily
saturate in the optical bands and specific procedures are needed to
try and correct for their blooming effects (see, e.g.,
\citealp{anderson08} for more details).  For every star thus
recovered, the multiple magnitude estimates obtained in each chip with
the same filter were homogenized by using \texttt{DAOMATCH} and
\texttt{DAOMASTER}, and their weighted mean and standard deviation
were finally adopted as the star magnitude and photometric error.

The instrumental magnitudes have been calibrated onto the VEGAMAG
photometric system by using the recipes and zero-points reported in
the {\emph HST} web-sites. The instrumental coordinates were first
corrected for geometric distortions by using the most updated ACS/WFC
Distortion Correction Tables (IDCTAB) provided in the dedicated web
page of the Space Telescope Science Institute.  Then, they were
reported to the absolute coordinate system ($\alpha,\delta$) as
defined by the World Coordinate System using the stars in common with
the publicly available Gaia DR2 catalog. The resulting astrometric
accuracy is typically $< 0.1$ mas.

\begin{figure*}[!t]
\centering
\includegraphics[width=1\textwidth]{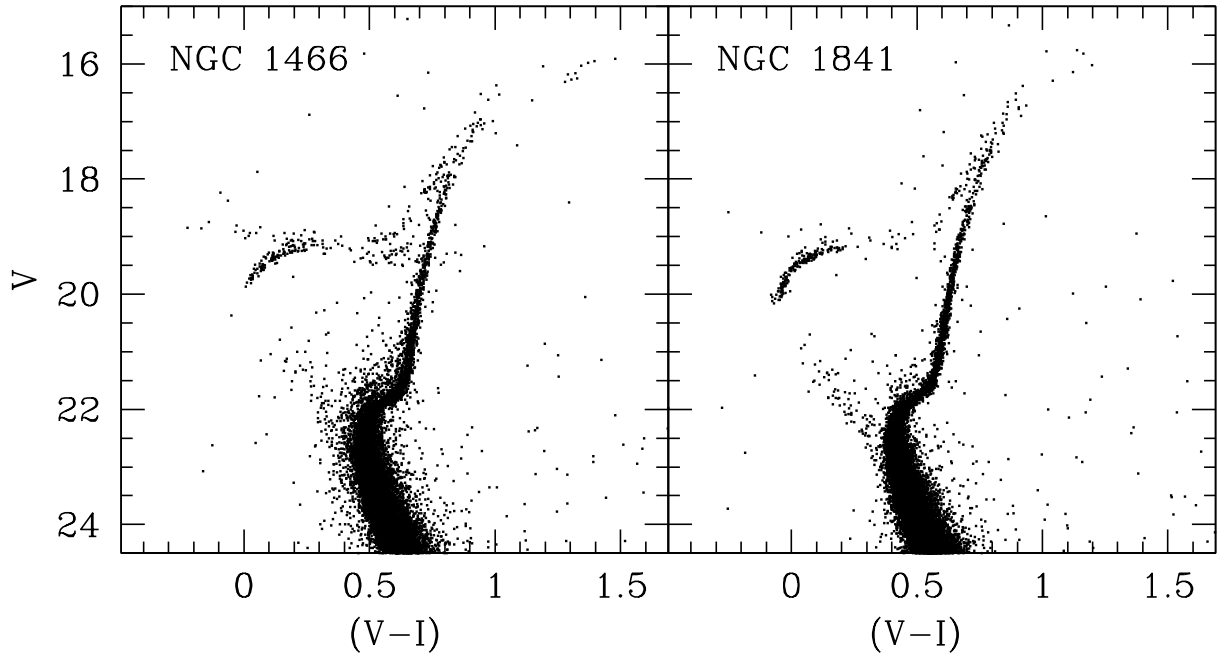}
\caption{Color-magnitude diagrams of NGC 1466 and NGC 1841 obtained
  from the analysis of the ACS/WFC images listed in Table
  \ref{tab_data}, following the procedure described in Section
  \ref{sec_data}. Throughout the paper $V$, and $I$ indicate the
  magnitudes obtained with the F606W and F814W filters, respectively.}
\label{cmd2}
\end{figure*}

\begin{figure*}[!t]
\centering
\includegraphics[width=1\textwidth]{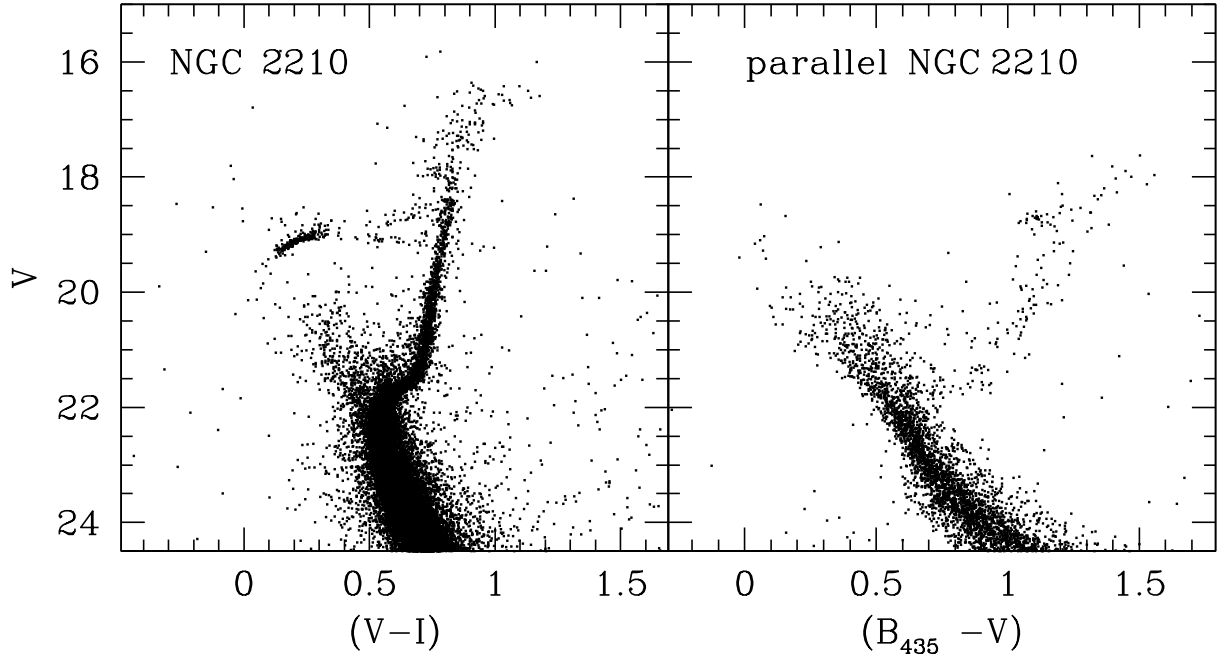}
\caption{CMDs obtained from the cluster pointing (left panel) and the
  parallel field (right panel) in the direction of NGC
  2210. Throughout the paper $B_{435}$ indicates the F435W magnitude.}
  \label{cmd2210}
\end{figure*}

\begin{figure*}[!h]
\centering
\includegraphics[width=1\textwidth]{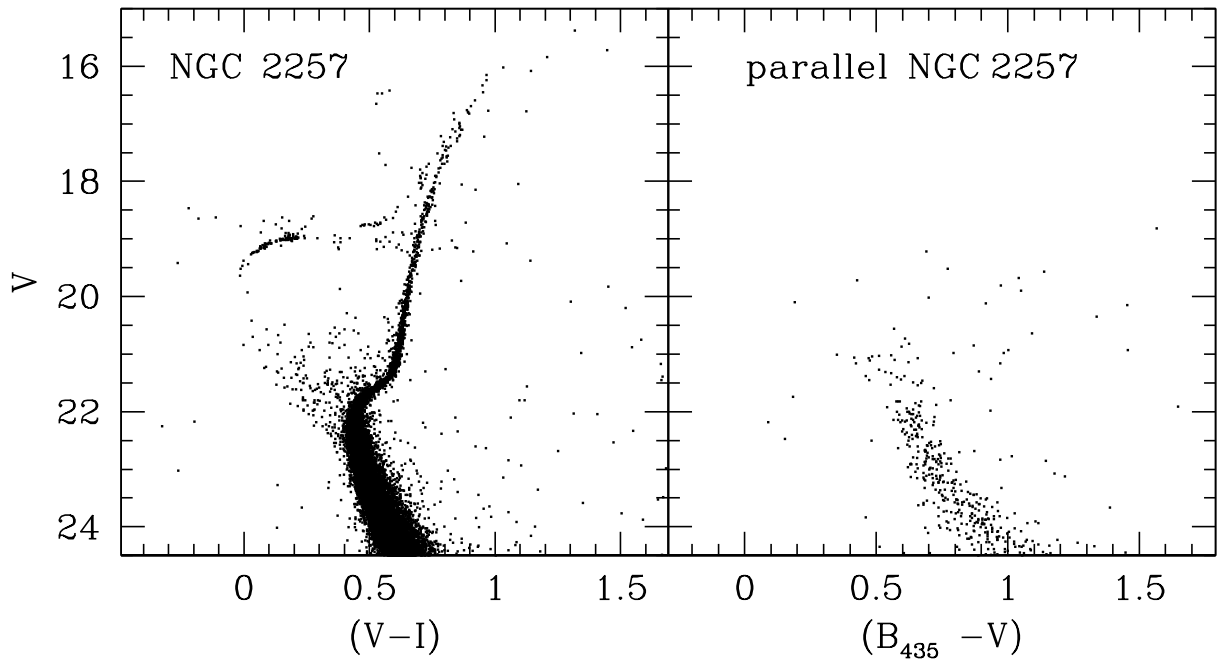}
\caption{As in Figure \ref{cmd2210}, but for NGC 2257.}
  \label{cmd2257}
\end{figure*}

\begin{figure*}[!h]
\centering
\includegraphics[width=1\textwidth]{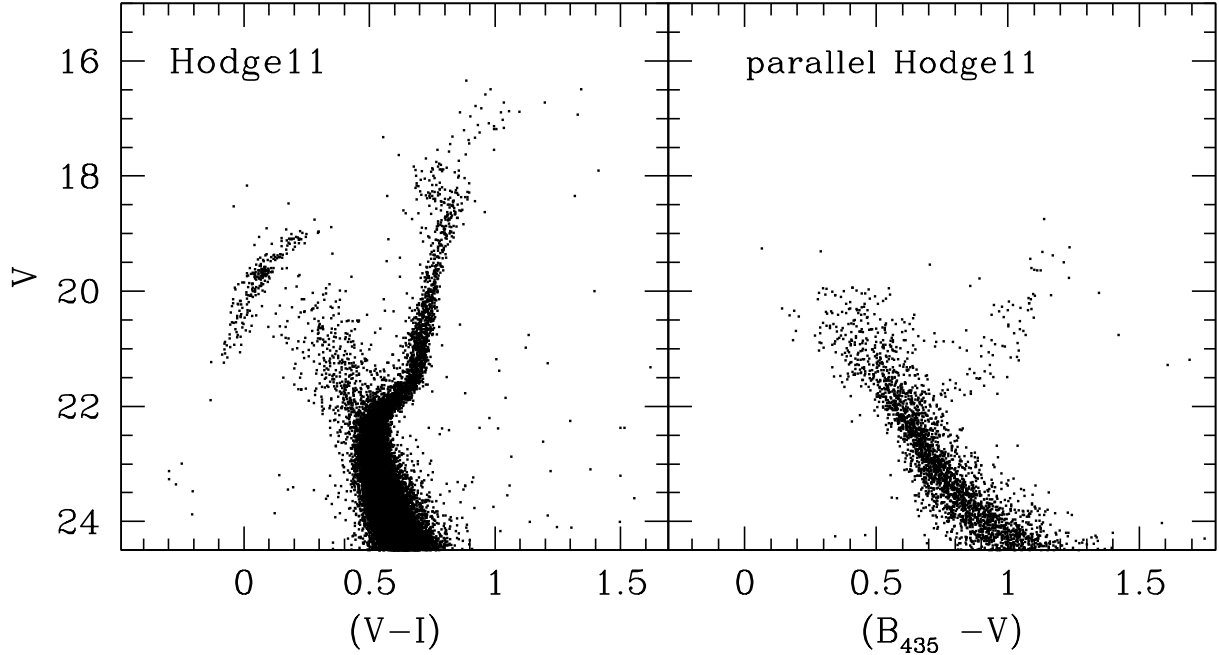}
\caption{As in Figure \ref{cmd2210}, but for Hodge 11.}
  \label{cmdhodge}
\end{figure*}

\begin{figure*}[!h]
\centering
\includegraphics[width=1\textwidth]{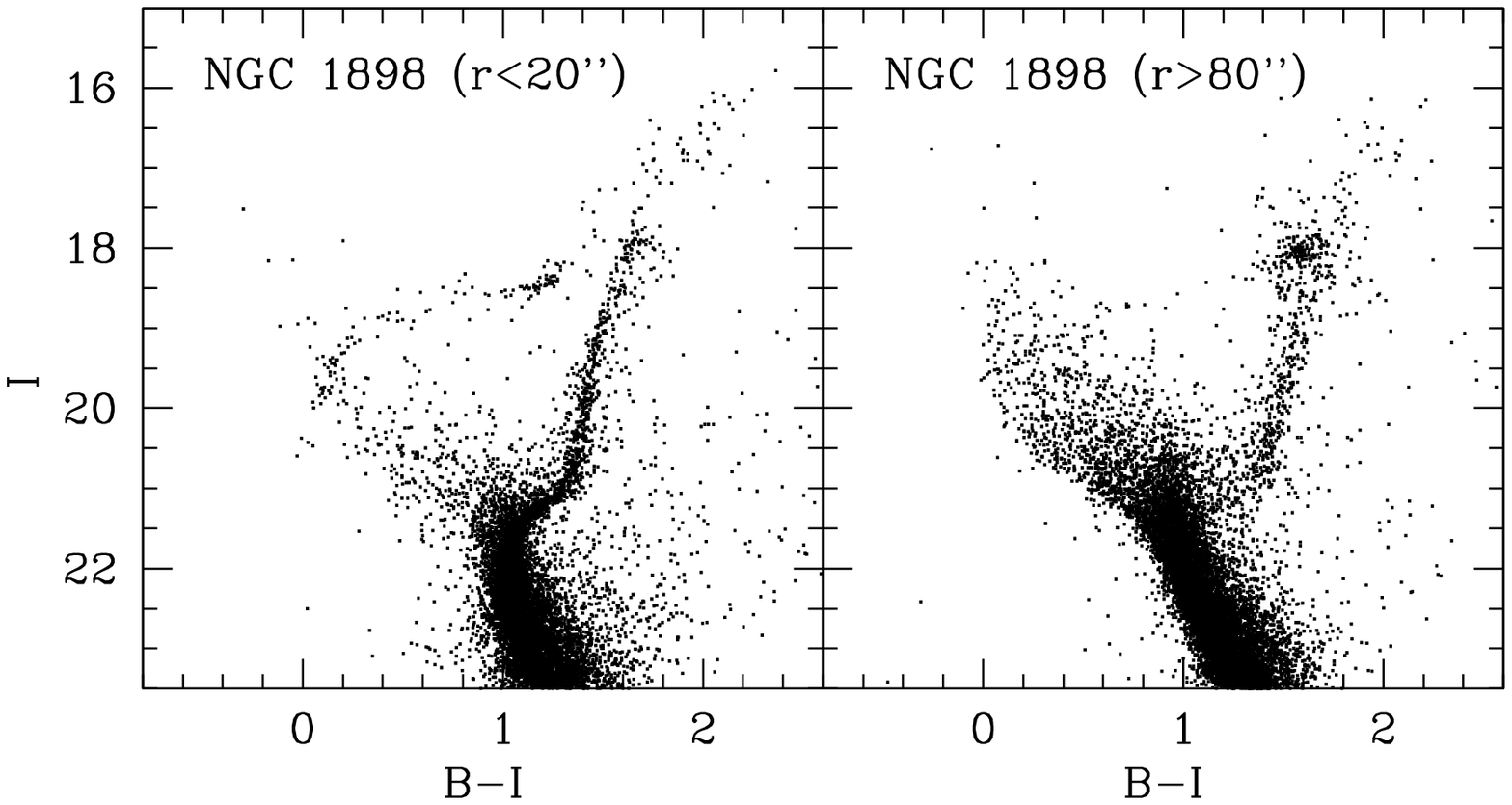}
\caption{CMDs of NGC 1898 obtained from all the stars measured within
  $20\arcsec$ and beyond $80\arcsec$ from the center of the system
  (right and left panels, respectively). While the cluster population
  is well visible at small radii, the contribution of the LMC field
  becomes dominant in the outskirts (see also Figure
  \ref{prof2}). Throughout the paper $B$ indicates the F475W
  magnitude.}
\label{cmd1898}
\end{figure*}

Figures \ref{cmd2}--\ref{cmd1898} show the color-magnitude diagrams
(CMDs) obtained from these data in the direction of all the program
clusters. As apparent, the cluster pointings are deep enough to nicely
trace all the evolutionary sequences, reaching $\sim 2$ magnitudes
below the main sequence turn-off (MS-TO) level. The CMDs obtained from
the parallel observations of NGC 1466 and NGC 1841 are very sparse,
with just 53 and 61 stars above $V=23$, respectively (and thus they
are not shown). Those obtained in the nearby fields of NGC 2210, NGC
2257 and Hodge11 are, instead, more populated (see right panels in
Figures \ref{cmd2210}--\ref{cmdhodge}), indicating that these three
systems are located in denser regions of the LMC. Indeed, the presence
of LMC field stars is well visible also in the corresponding cluster
CMDs, especially for NGC 2210 and Hodge11 (Figure \ref{cmd2210} and
\ref{cmdhodge}), as a prominent extension of the cluster MS at
magnitudes brighter than the MS-TO point.  In Figure \ref{cmd1898} we
show the CMD of NGC 1898, plotting separately the stars measured
within $20\arcsec$ (illustrating the distribution of the cluster
population; left panel), and those beyond $80\arcsec$ from the center,
which are dominated by the LMC field population (right panel).

\section{Center of gravity}
\label{sec_center}
For a proper determination of the projected star density profiles, it
is first necessary to accurately identify the center of each system.
As discussed in previous papers, we were among the first groups in
promoting the adoption of the gravitational center derived from star
counts ($C_{\mathrm grav}$), in place of the location of the surface
brightness peak, as the best proxy of the cluster center
\citep{montegriffo+95}. This is to avoid any possible bias induced by
the presence of a few bright star, which would significantly alter the
location of the surface brightness maximum.

The determination of $C_{\mathrm grav}$ requires the selection of a
sample of resolved stars large enough (a few thousand objects) to
guarantee high statistics, while avoiding spurious effects due to
photometric incompleteness (which increases for increasing magnitude
and decreasing radial distance). By means of artificial star
experiments (see, e.g., \citealp{bellazzini02, beccari10, ema15,
  sollima+17}) we estimated that the photometric completeness is above
80\% at all radii for magnitudes $\sim 0.5-1.5$ below the MS-TO level
in the less dense clusters. The magnitude limit for comparable
completeness levels is $\sim 0.8$ above the MS-TO for NGC 2210, which
is the most concentrated system.  These thresholds have been used to
select a ``representative sample'' of stars in each cluster for the
determination of $C_{\mathrm grav}$.  According to what discussed
above, the adopted selection includes a fraction of LMC field stars.
However, within the small sky area ($\sim 200\arcsec\times
200\arcsec$) covered by the ACS/WFC in the direction of each system,
field stars are expected to have a uniform radial distribution with
respect to the cluster center and they thus introduce no biases in the
identification of $C_{\mathrm grav}$.  We then followed the iterative
procedure already adopted in previous works \citep[see,
  e.g.,][]{lanzoni+07a,lanzoni+07b,lanzoni+10, miocchi+13}: we
selected all the stars belonging to the ``representative sample'' and
falling within a circle of radius $r$ from a first-guess center; the
average of their coordinates projected on the plane of the sky ($x$
and $y$) provided us with a new guess value for the center, from which
we iteratively repeated the procedure until convergence.  We assumed
that convergence is reached when ten consecutive iterations yield
values of the cluster center that differ by less than $0.01\arcsec$
from each other.  As first-guess center, we adopted the values quoted
in Table 4 of MG03. The optimal value for the search radius $r$ cannot
be known a priori, but it must exceed the cluster core radius to be
sensitive to the portion of the profile where the slope changes and
the density is no more uniform (see \citealt{miocchi+13}). On the
other hand, too large radii result in a reduced sensitivity to the
central concentration.  Hence, reasonable values of $r$ typically
range between a few arcseconds and a few dozens of arcseconds larger
than the cluster core radius (taken from Mv05, in this case),
depending on the structure of the system. Of course, adopting
different values of $r$ (and/or different magnitude cuts for the
selection of the stellar sample) does not exactly yield to the same
average position of the stars. Hence, in order to estimate both
$C_{\rm grav}$ and its uncertainty, we repeated the procedure by
assuming different values of $r$ and different magnitude limits,
finding a different ``convergence center'' for every pair of these
parameters. The average of the obtained values has been finally
assumed as the gravitational center of the cluster, and their
dispersion is adopted as uncertainty. For all the target clusters, we
considered at least three values of $r$ (following the criteria
discussed above) and three limiting magnitudes. These latter typically
range between the 80\% completeness threshold, and $\sim 0.5-1$
magnitude brighter, thus guaranteeing that at least a few hundreds of
stars are always included within the search radius, which is necessary
to make the average of their position statistically significant.

\begin{figure*}[!t]
\centering
\includegraphics[width=1\textwidth]{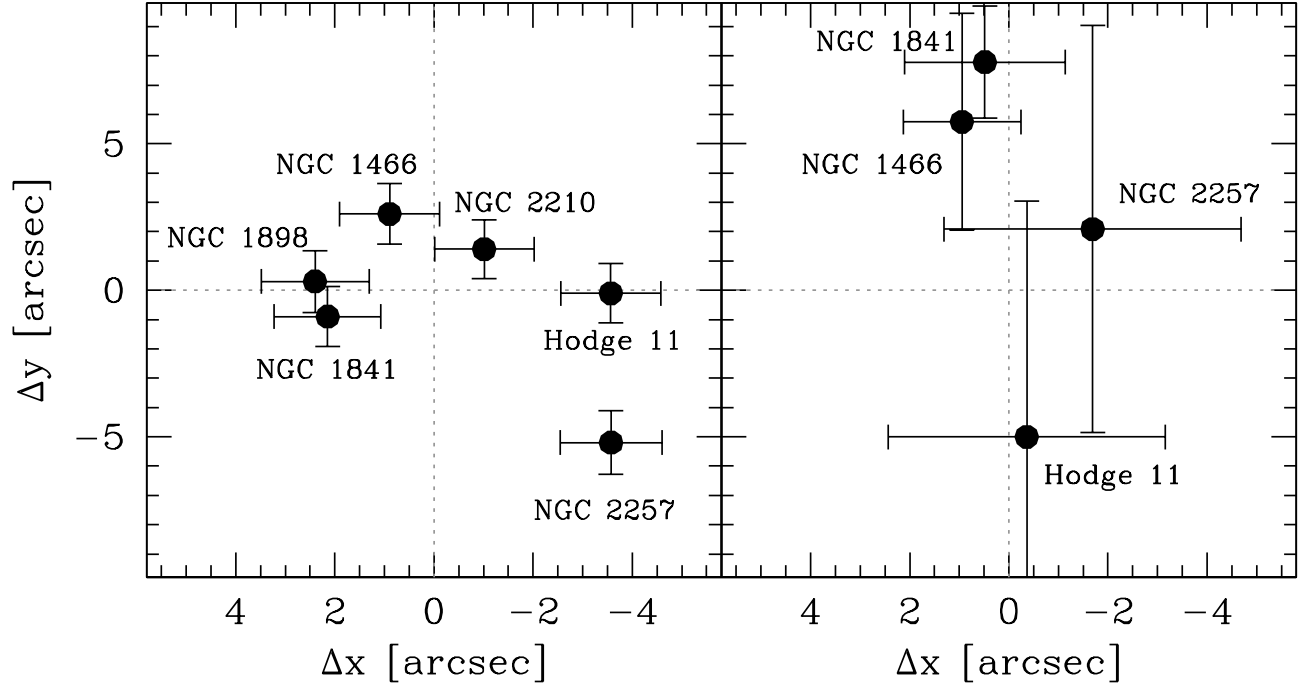}
\caption{{\emph Left panel:} Distance on the plane of the sky, along
  the RA and the Dec directions ($\Delta x$ and $\Delta y$,
  respectively), between the cluster centers estimated by MG03 (see
  their Table 4) and those determined in the present work (Table
  \ref{tab_centers}). {\emph Right panel:} The same as in the left
  panel for the four GCs in common with S18 (see their Table 2).}
\label{centers}
\end{figure*}

The values derived for the six program clusters are listed in Table
\ref{tab_centers} and compared to those quoted by MG03 and
\citet[][hereafter S18]{sun18} in the left and right panels of Figure
\ref{centers}, respectively.  The figure shows the distance on the
plane of the sky between the literature centers and those determined
in the present work for the GCs in common. The errorbars have been
computed as the square root of the quadratic sum of the quoted
uncertainties.  Given the large uncertainties quoted in S18, the two
determinations are consistent within 1-1.5 $\sigma$, with the only
exception of NGC 1841 showing a $\Delta y$ discrepancy larger than 3
$\sigma$.  With respect to the centers quoted in MG03, we find
differences as large a few arcseconds along either the right
ascension, or the declination directions, or both.  The center showing
the largest discrepancy is that of NGC 2257, that MG03 locate $\sim
6\arcsec$ ($\sim 1.5$ pc at the distance of the LMC, 50 kpc) far away
from ours, in the south-west direction.  The origin of the
disagreements is likely ascribable to the different methods adopted in
these studies, with S18 using a criterion based on the maximum spatial
density determined through a two-dimensional Gaussian kernel density
estimator, and MG03 referring to the surface brightness peak measured
in {\emph HST}/WFPC2 images after corrections for the biasing effect
of the brightest cluster stars.  In any case, such non negligible
differences can affect the derived shape of the star density profile
and, in general, the study of the radial distribution of all stellar
populations, especially for the most concentrated systems (see Section
\ref{sec_discuss}).

\begin{table}[t!]
\caption{Center of gravity}
\begin{center}
{\begin{tabular}{lccc} \hline\hline Cluster & RA & Dec & $\sigma_{\rm rms}$ \\ \hline \hline
NGC 1466 & $03^{\rm h}$ $44^{\rm m}$ $32.71^{\rm s}$ & $-71\arcdeg$ $40\arcmin$ $15.6\arcsec$ & $0.3\arcsec$ \\
NGC 1841 & $04^{\rm h}$ $45^{\rm m}$ $22.53^{\rm s}$ & $-83\arcdeg$ $59\arcmin$ $55.1\arcsec$ & $0.4\arcsec$ \\
NGC 1898 & $05^{\rm h}$ $16^{\rm m}$ $41.94^{\rm s}$ & $-69\arcdeg$ $39\arcmin$ $25.3\arcsec$ & $0.5\arcsec$ \\
NGC 2210 & $06^{\rm h}$ $11^{\rm m}$ $31.69^{\rm s}$ & $-69\arcdeg$ $07\arcmin$ $18.4\arcsec$ & $0.1\arcsec$ \\
NGC 2257 & $06^{\rm h}$ $30^{\rm m}$ $12.65^{\rm s}$ & $-64\arcdeg$ $19\arcmin$ $36.8\arcsec$ & $0.5\arcsec$ \\
Hodge 11 & $06^{\rm h}$ $14^{\rm m}$ $22.99^{\rm s}$ & $-69\arcdeg$ $50\arcmin$ $49.9\arcsec$ & $0.2\arcsec$ \\
\hline
\hline
\end{tabular}}
\label{tab_centers}
\end{center}
\tablecomments{Coordinates of the gravity centers determined in the
  present work for the program clusters. The estimated uncertainties
  ($\sigma_{\rm rms}$) are listed in column 4.}
\end{table}

\section{Star count density profiles}
\label{sec_dens}
To determine the projected density profile of each cluster we used the
same ``representative samples'' of stars discussed above, and number
counts corrected following the completeness curves obtained at
different radial distances from the center. We divided the field of
view observed in the direction of each cluster in a number of
concentric annuli (typically 10-20) centered on $C_{\rm grav}$ and
split in (typically four) sub-sectors.  We then counted the
completeness-corrected number of stars lying within each sub-sector
and divided it by the sub-sector area. The stellar density in each
annulus was finally obtained as the average of the sub-sector
densities, and the standard deviation among the sub-sectors densities
was adopted as error.  To also take into account the completeness
uncertainty, we repeated the procedure two more times, by correcting
the number counts according to the curves obtained after the
subtraction and the addition of the completeness uncertainties to the
radial completeness curves obtained from the artificial star
experiments. The final error bar of the projected density at every
radial bin is thus assumed to span the entire range covered by the
errors of the three resulting profiles.  For all the clusters but NGC
1898, we performed the analysis also on the nearby fields sampled by
the parallel observations, using the same magnitude thresholds adopted
for the cluster populations, taking advantage of the fact that the
cluster and parallel pointings have the $V$ filter in common.

\begin{figure}[!t]
\centering
\hbox{
\includegraphics[height=9truecm,width=9truecm]{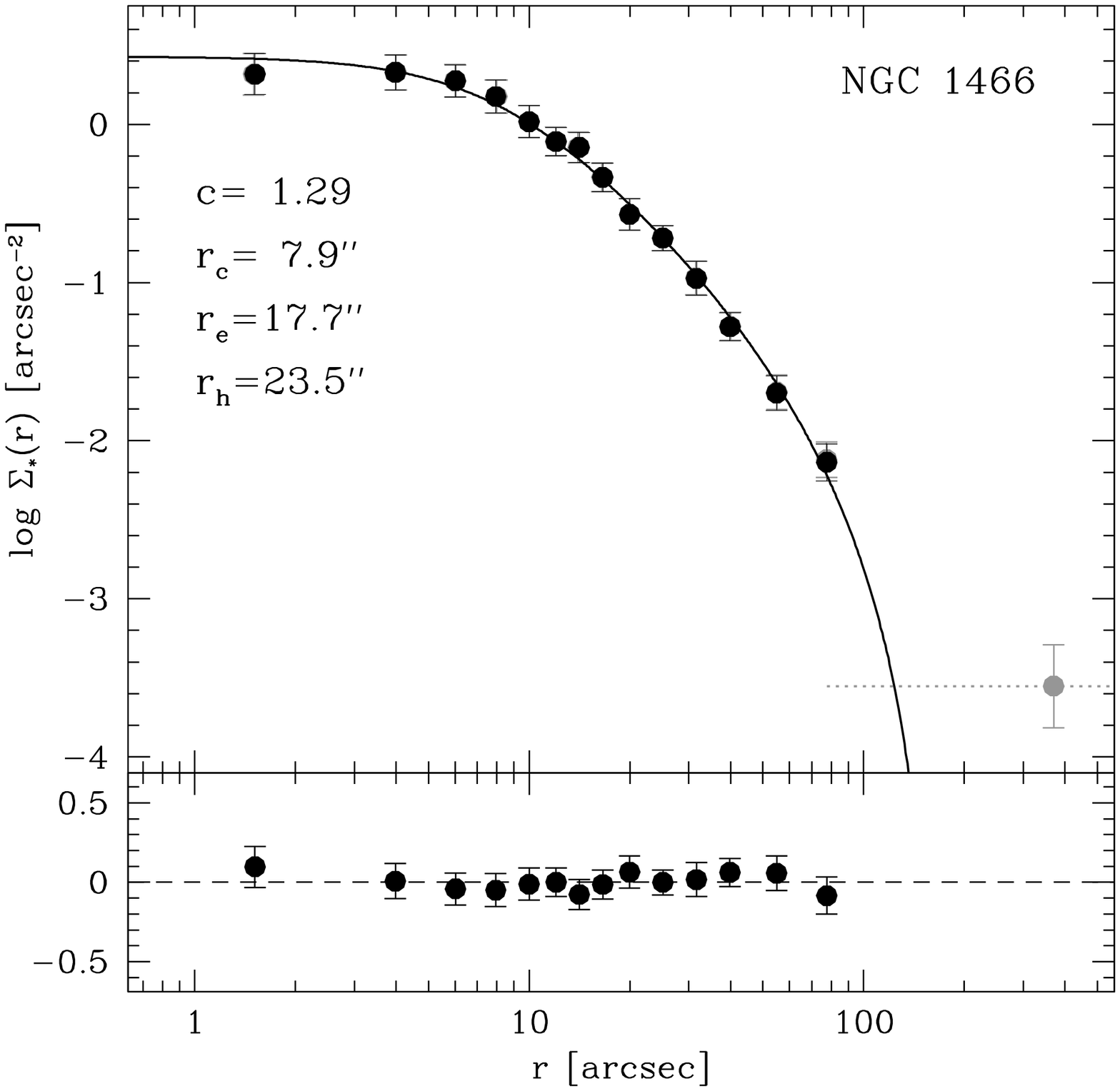}
\includegraphics[height=9truecm,width=9truecm]{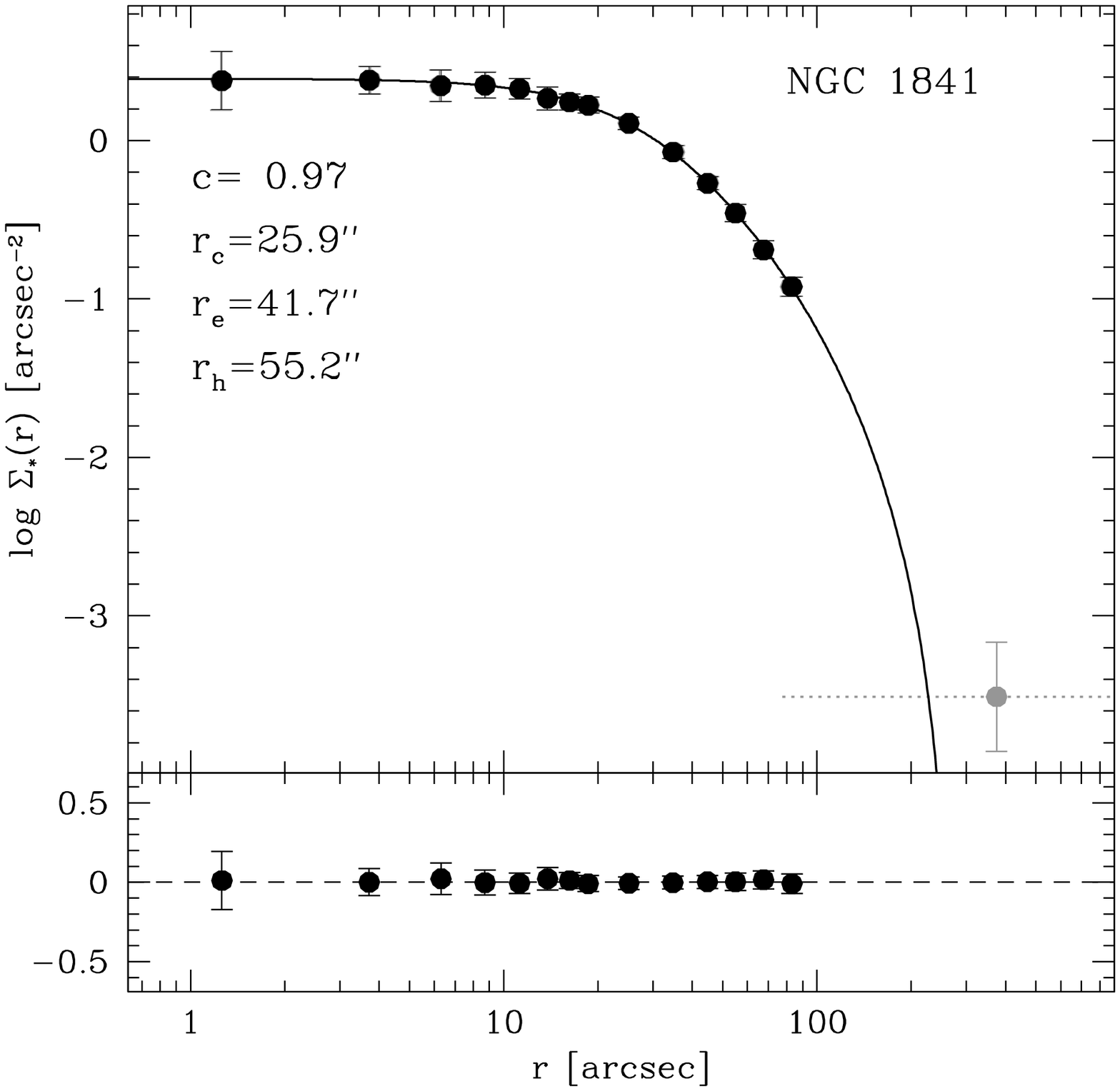}
}
\caption{Projected density profiles of NGC 1466 (left) and NGC 1841
  (right) obtained in the present work from resolved star counts.  The
  black circles correspond to the cluster density profile obtained
  after subtraction of the LMC field contribution (grey circles and
  dotted lines). The black lines show the best-fit King model
  profiles, with the corresponding values of the concentration
  parameter ($c$) and a few characteristic scale-lengths (in
  arcseconds) labelled. The residuals between the model and the
  observations are plotted in the bottom portion of each panel.}
\label{prof1}
\end{figure}

\begin{figure}[!t]
\centering
\hbox{
\includegraphics[height=9truecm,width=9truecm]{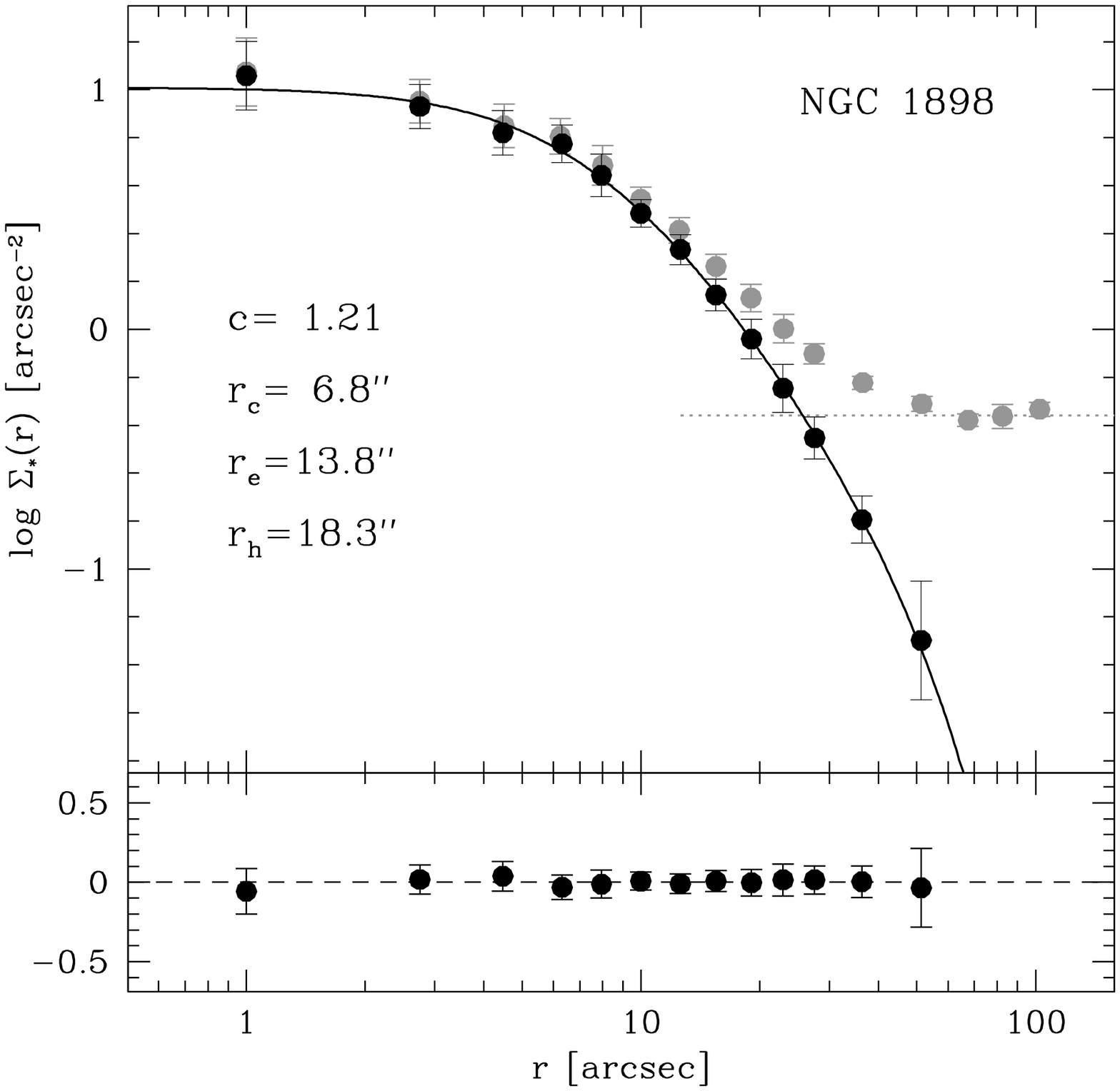}
\includegraphics[height=9truecm,width=9truecm]{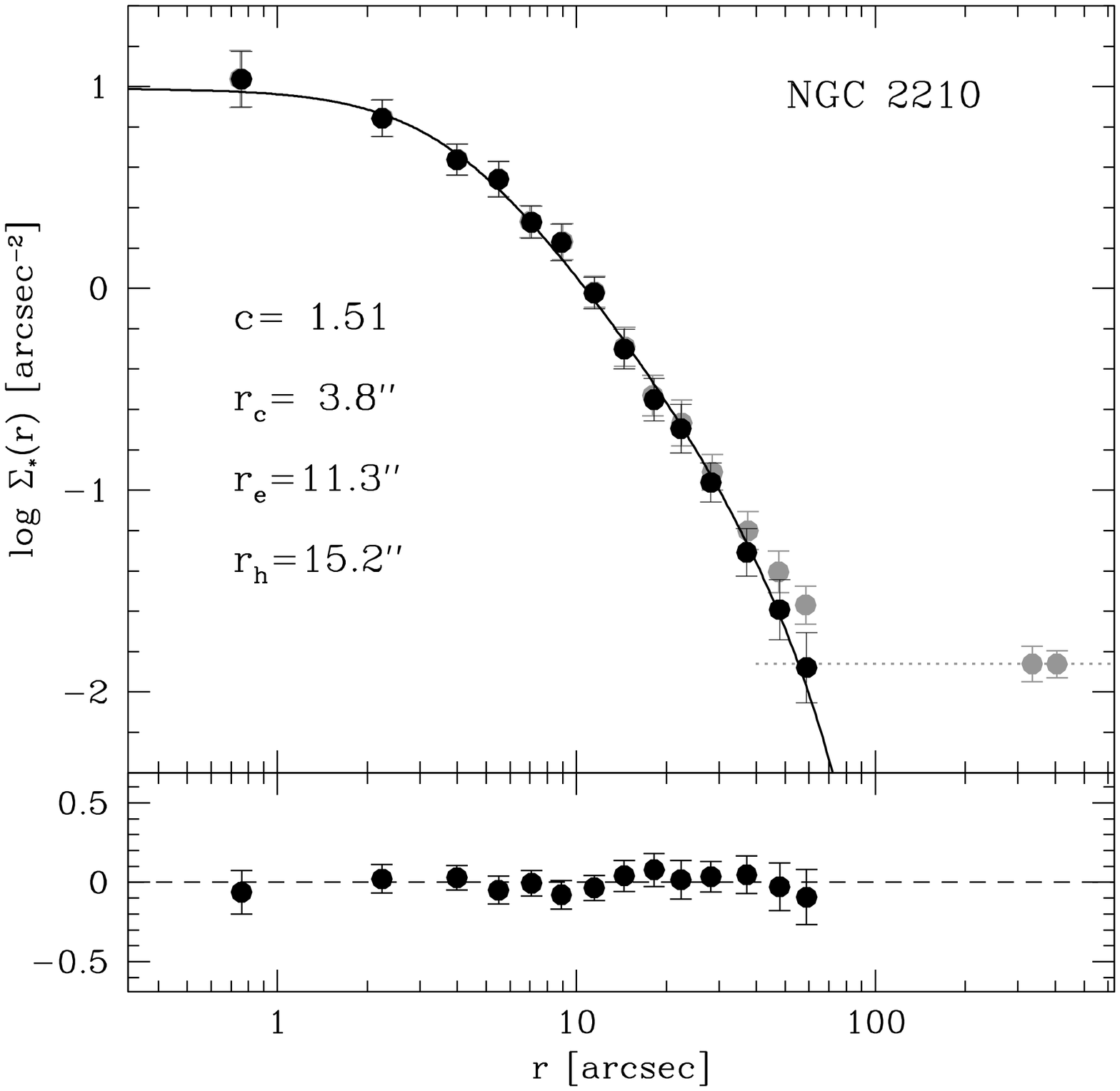}
}
\caption{As in Figure \ref{prof1}, but for NGC 1898 and NGC 2210.}
\label{prof2}
\end{figure}

\begin{figure}[!t]
\centering
\hbox{
\includegraphics[height=9truecm,width=9truecm]{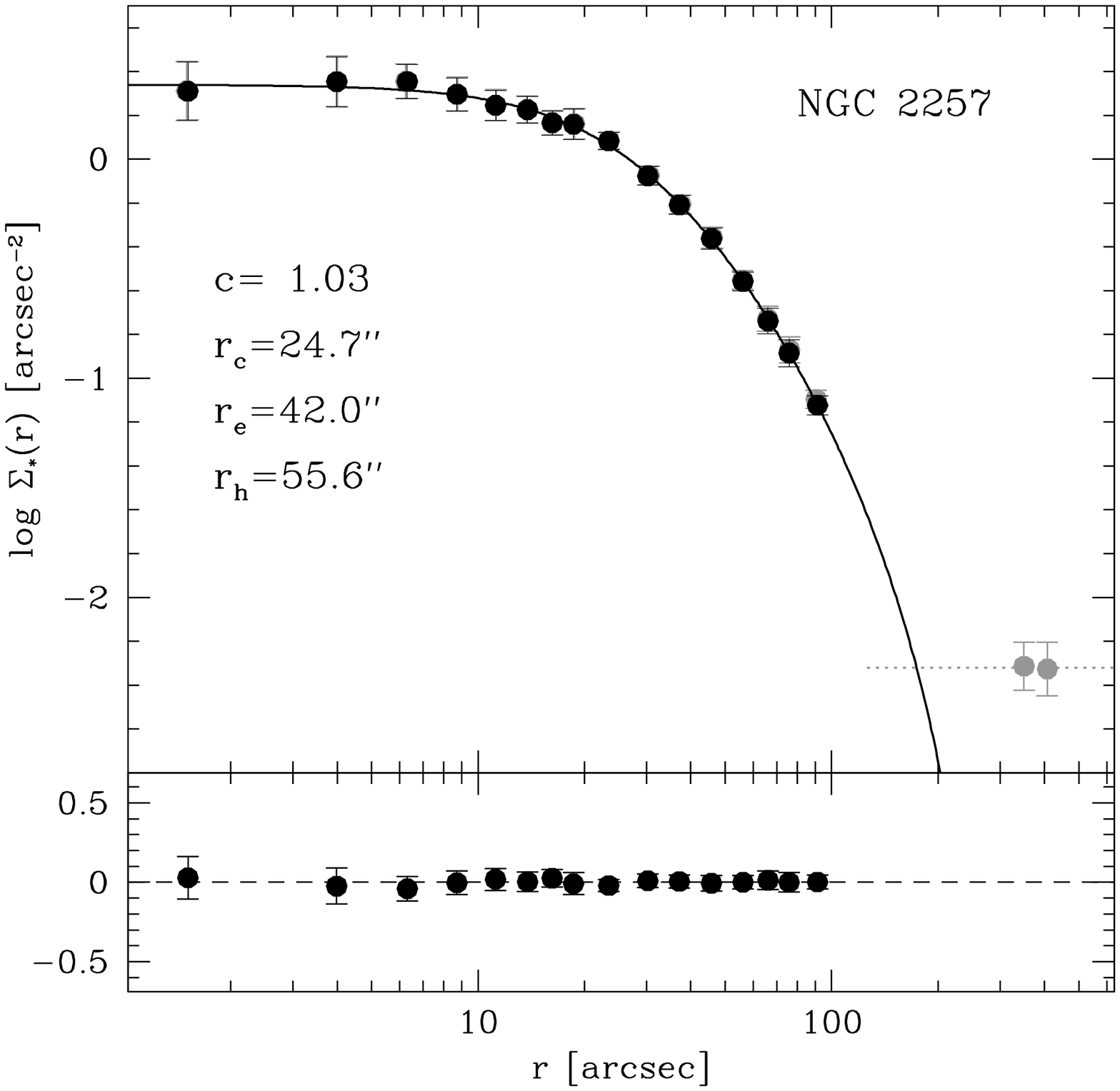}
\includegraphics[height=9truecm,width=9truecm]{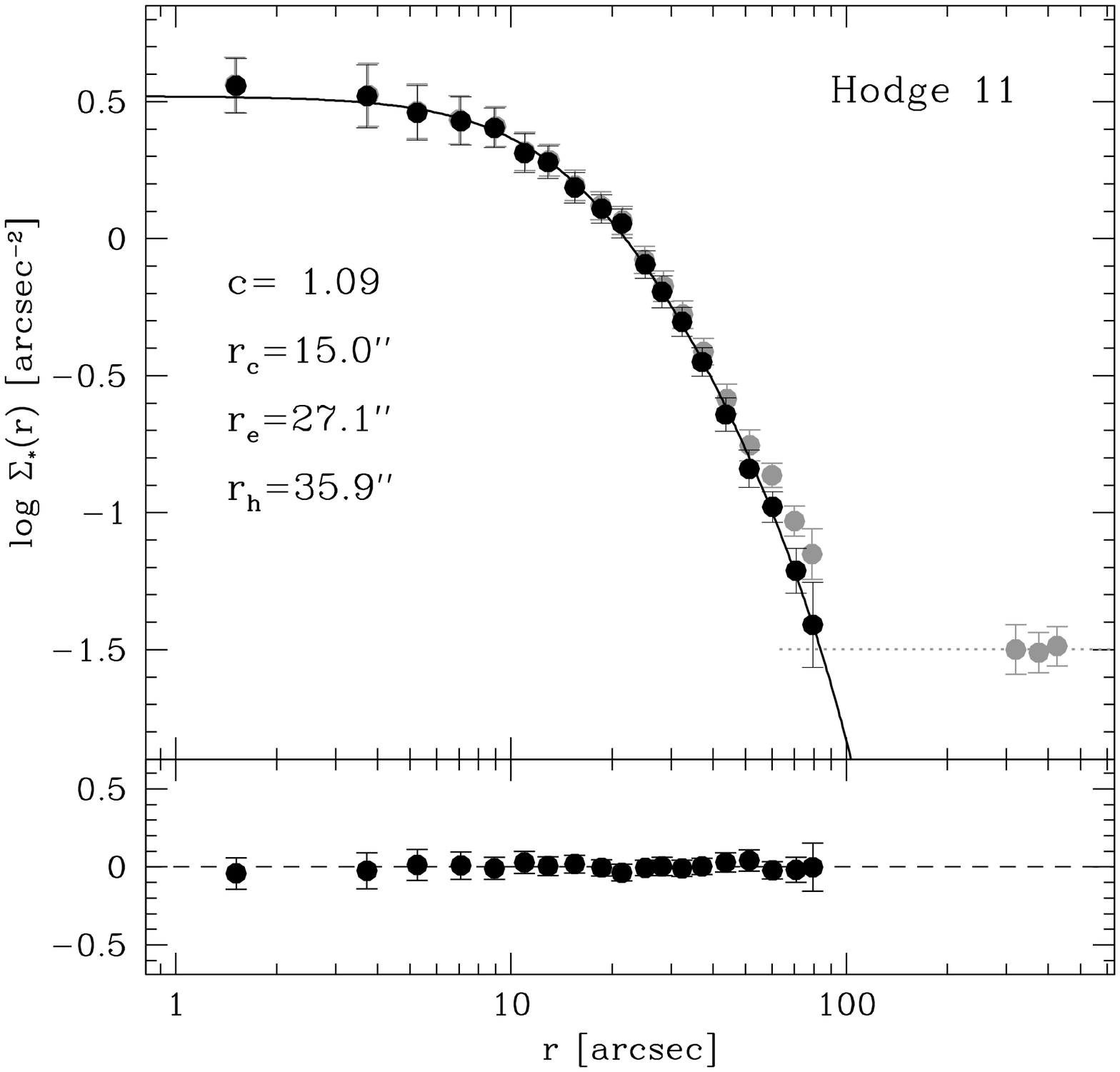}
}
\caption{As in Figure \ref{prof1}, but for NGC 2257 and Hodge 11.}
\label{prof3}
\end{figure}

The resulting stellar density profiles $\Sigma_*(r)$, in units of
number of stars per square arcsecond, are shown in Figures
\ref{prof1}--\ref{prof3} for the six program clusters (grey circles).
The constant values observed at large radii correspond to the LMC
field density. As expected from Figure \ref{cmd1898}, the field
contribution is particularly high in the case of NGC 1898.  For each
cluster, we thus averaged the densities observed in the external
plateau\footnote{For NGC 1466 and NGC 1841, the sparseness of the
  parallel field CMDs allows us to measure just one point, that we
  adopt as LMC field density level.}  and subtracted this value
(short-dashed lines in Figures \ref{prof1}--\ref{prof3}) from the
observed density distribution.  The true cluster density profile,
obtained after subtraction of the LMC background, is finally shown
with black circles in the figures.  As apparent, in the inner regions
(where the cluster density is much larger than the field one), the
background-subtracted profile remains unchanged with respect to the
observed distribution. This holds for the entire radial range sampled
in NGC 1466, NGC 1841 and NGC 2257, because the field density is
orders of magnitude lower than the outermost value measured from the
cluster pointings. For the other clusters, instead, the background
subtraction significantly reduces the density in the outer portion of
the profile, where the LMC contribution become increasingly more
important. Indeed, after background subtraction, the external cluster
density can be significantly lower than the observed LMC field
level. Hence, an accurate measure of the background contribution in
the direction of each cluster is crucial for a reliable determination
of the outermost portion (and thus the overall shape) of the density
profile.

\section{Models}
\label{sec_models}
By construction, the ``reference samples'' used to build the density
profiles include approximately equal-mass stars, In fact, the
difference in mass between objects at the MS-TO level or just below
it, and those evolving in any post-MS evolutionary phases is very
small (within a few 0.01 $M_\odot$). Hence, to determine the physical
parameters of the program clusters we used single-mass, spherical and
isotropic \cite{king66} models.  This is a \emph{single}-parameter
family, where the shape of the density profile is uniquely determined
by the dimensionless parameter $W_0$, which is proportional to the
central gravitational potential of the system.  These models are
characterized by a constant projected density in the innermost region
(corresponding to the so-called cluster ``core'') and a decreasing
behavior outwards.  In practice, the higher is $W_0$, the smaller is
the cluster core with respect to the overall size of the
system. Indeed, there is a one-to-one relation between $W_0$ and the
concentration parameter $c$, defined as $c\equiv\log(r_t/r_0)$, where
$r_t$ is the tidal (or truncation) radius of the system and $r_0 $ is
the model radial scale (the so called ``King radius''). Hence, the
King models are often equivalently parametrized in terms of either
$W_0$, or $c$.

To determine the best-fit model for each of the surveyed clusters, we
compared the background-subtracted surface density profiles with the
King model family, leaving $W_0$ to vary from 4.0 to 12.0 in steps of
0.05 (the corresponding concentration parameter $c$ varies between
0.84 and 2.74).  For every explored value $W_{0,i}$, we determined the
scale parameter $r_{0,i}$ and the central surface density
$\Sigma_{*0,i}$ providing the minimum reduced $\chi^2$ of the
residuals between the observed and the model profiles ($\chi^2_{{\rm
    min}, i}$).  The solution corresponding to the lowest value of the
stored $\chi^2_{{\rm min}, i}$ ($\chi^2_{\rm best}$) is finally
adopted as the best-fit model. Beside the best-fit values of $W_0$,
$r_0$ and $\Sigma_{*0}$, this also provides several characteristic
scale-lengths useful for the physical description of the cluster
structure, as (see, e.g., \citealt{miocchi+13}): the ``core radius''
($r_c$), which is operatively defined as the radius at which the
\emph{projected} stellar density $\Sigma_*(r)$ drops to half of its
central value (in other studies the surface brightness is considered
instead of $\Sigma_*$; this scale-length is not equivalent to the King
radius $r_0$, although they become increasingly similar for increasing
$W_0$ or $c$); the ``half-mass radius'' ($r_h$), which is the radius
of the sphere containing half the total cluster mass (of course,
observations does not provide this 3-dimensional quantity); the
``effective radius'' ($r_e$), which is the radius of the circle that,
\emph{in projection}, includes half the total counted stars (in
studies using the surface brightness instead of number counts, this is
the radius containing half the total luminosity in projection).

The best-fit models are shown as black lines in Figures
\ref{prof1}-\ref{prof3}, and their residuals with respect to the
observed profiles are plotted in the bottom panels. Table
\ref{tab_params} lists the best-fit parameters together with their
uncertainties estimated from the maximum variations of each parameter
within the subset of models that provide a $\chi^2_{{\rm min},i} \le
\chi^2_{\rm best} +1$ (see Mv05, \citealp{miocchi+13}). For each scale
radius we also quote the value in parsec, computed by assuming a
distance of 50 kpc for the LMC \citep{pietrzynski13}.

\begin{table}[t!]
\caption{Structural parameters}
\begin{center}
{\begin{tabular}{lrrrrrrrrr}
\hline\hline
Cluster & $c$ & $r_c$ & $r_e$ & $r_h$ & $r_t$ & $r_c$ & $r_e$ & $r_h$ & $r_t$\\
        &   & [$\arcsec$] & [$\arcsec$] & [$\arcsec$] & [$\arcsec$] & [pc]        & [pc]        & [pc]        & [pc] \\
\hline \hline\\[-10pt]
NGC 1466 & $1.29^{+0.09}_{-0.07}$ & $ 7.9^{+0.8}_{-0.8}$    & $17.7^{+0.2}_{-0.1}$   & $23.5^{+0.3}_{-0.1}$    & $167.0^{+16.7}_{-10.4}$ & $1.9^{+0.2}_{-0.2}$  & $4.3^{+0.1}_{-0.1}$  & $5.7^{+0.1}_{-0.1}$  & $40.5^{+4.0}_{-2.5}$\\[3pt]
\hline\\[-10pt]
NGC 1841 & $0.97^{+0.12}_{-0.13}$ & $ 25.9^{+1.6}_{-0.9}$   & $41.7^{+3.4}_{-1.9}$   & $55.2^{+4.7}_{-2.7}$    & $278.4^{+66.6}_{-47.2}$ & $6.3^{+0.4}_{-0.2}$  & $10.1^{+0.8}_{-0.5}$ & $13.4^{+1.1}_{-0.7}$ & $67.5^{+16.1}_{-11.4}$ \\[3pt] 
\hline\\[-10pt]
NGC 1898 & $1.21^{+0.13}_{-0.11}$ & $ 6.8^{+0.6}_{-0.6}$    & $13.8^{+0.7}_{-0.4}$   & $18.3^{+1.0}_{-0.6}$    & $119.3^{+23.3}_{-17.2}$ & $1.6^{+0.1}_{-0.1}$ & $3.3^{+0.2}_{-0.1}$  & $4.4^{+0.2}_{-0.1}$  & $28.9^{+5.6}_{-4.2}$ \\[3pt]
\hline\\[-10pt]
NGC 2210 & $1.51^{+0.09}_{-0.07}$ & $ 3.8^{+0.4}_{-0.4}$    & $11.3^{+0.3}_{-0.2}$   & $15.2^{+0.5}_{-0.3}$    & $130.3^{+12.0}_{-7.5}$ &  $0.9^{+0.1}_{-0.1}$ & $2.7^{+0.1}_{-0.1}$  & $3.7^{+0.1}_{-0.1}$  & $31.6^{+2.9}_{-1.8}$ \\[3pt]
\hline\\[-10pt]
NGC 2257 & $1.03^{+0.11}_{-0.10}$ & $24.7^{+1.4}_{-1.4}$    & $42.0^{+2.0}_{-1.3}$   & $55.6^{+2.7}_{-1.9}$    & $299.4^{+53.3}_{-39.1}$ & $6.0^{+0.3}_{-0.3}$ & $10.2^{+0.5}_{-0.3}$ & $13.5^{+0.7}_{-0.5}$ & $72.6^{+12.9}_{-9.5}$ \\[3pt] 
\hline\\[-10pt]
Hodge 11 & $1.09^{+0.10}_{-0.07}$ & $15.0^{+0.8}_{-1.0}$    & $27.1^{+1.0}_{-0.5}$   & $35.9^{+1.4}_{-0.7}$    & $207.0^{+32.4}_{-19.4}$ & $3.6^{+0.2}_{-0.2}$ & $6.6^{+0.2}_{-0.1}$  & $8.7^{+0.3}_{-0.2}$  & $50.2^{+7.9}_{-4.7}$ \\[3pt]
\hline
\hline
\end{tabular}}
\label{tab_params}
\end{center}
\tablecomments{Concentration parameter ($c$), core radius ($r_c$),
  effective radius ($r_e$), half-mass radius ($r_h$) and tidal radius
  ($r_t$) obtained for the surveyed clusters from the best-fit King
  models to the observed density profiles. For every cluster, each
  radial scale is quoted both in arcseconds (columns 3--6) and in
  parsec (7--10), assuming a distance of 50 kpc.}
\end{table}

\begin{figure*}[!t]
\centering
\includegraphics[width=1\textwidth]{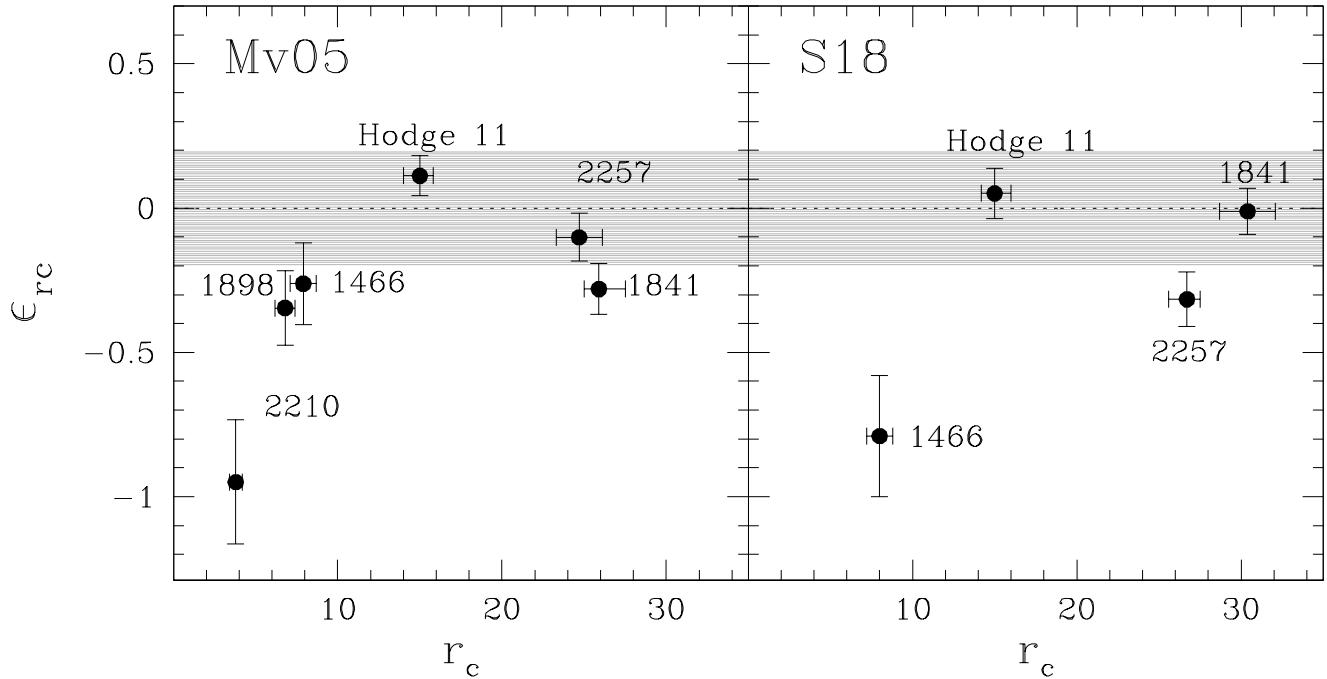}
\caption{Comparison between the values of the King core radii
  determined here, and those quoted in Mv05 (left) and in S18 (right).
  The x-axis shows the value of $r_c$, in arcseconds, estimated in the
  present work. The y-axis corresponds to the relative difference
  $\epsilon_{rc} = (r_c -r_{c, \rm let})/r$, where where $r_c$ is the
  value here derived, while $r_{c, \rm let}$ is the one quoted in Mv05
  or S18.}
\label{cfr_rc}
\end{figure*}

\begin{figure*}[!t]
\centering
\includegraphics[width=1\textwidth]{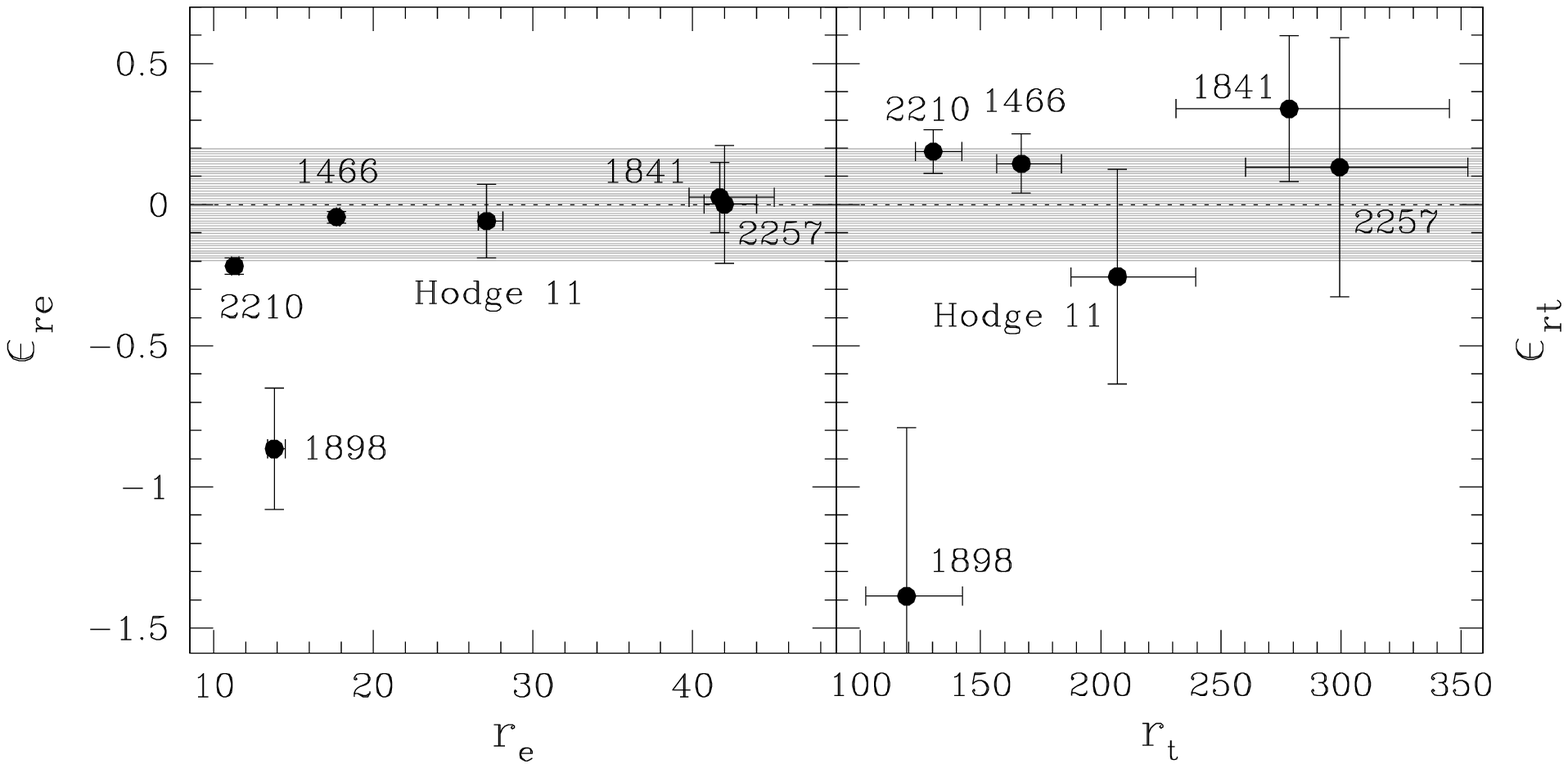}
\caption{Comparison between the values of the King effective and tidal
  radii (left and right panels, respectively) determined here and in
  Mv05.  The x-axis shows the values of $r_e$ and $r_t$, in
  arcseconds, estimated in the present work. The y-axis corresponds to
  the relative difference defined as in Figure \ref{cfr_rc}, but for
  the effective and tidal radii, instead of $r_c$.}
\label{cfr_rert}
\end{figure*}

\section{Discussion and conclusions}
\label{sec_discuss}
Figures \ref{prof1}-\ref{prof3} clearly show that the observed density
profiles of all the investigated clusters are very well reproduced by
King models. These span a large range of core radii, from just $\sim
4\arcsec$ (1 pc) for NGC 2210, up to $30\arcsec$ (7 pc) for NGC 1841,
which is also the system with the lowest concentration parameter $c$.

The comparison between the scale-radii estimated in this work and
those derived by Mv05 and S18 is shown in Figures \ref{cfr_rc} and
\ref{cfr_rert}.  Mv05 used the observed surface brightness (instead of
number count) profiles. S18 modeled the observed density distributions
with \citet{EFF} models, but then listed the corresponding values of
the King core radius in their Table 2. To transform into arcseconds
the values that Mv05 quote in parsec we used the LMC distance adopted
in their paper (50.1 kpc). The conversion of the S18 values is done by
assuming the distance moduli listed in their Table 3.  For each scale
radius, we plot the relative difference $\epsilon_{rx} = (r_x -r_{x,
  \rm let})/r_x$, where $r_x$ is the value of $r_c$, $r_e$ or $r_t$
here derived, and $r_{x, \rm let}$ is the corresponding one quoted
either in Mv05, or in S18 (for $r_c$ only).  In terms of $r_c$, the
comparison with the Mv05 values shows a good agreement for the three
largest systems, while a notable discrepancy is found for NGC 2210,
which is the most compact cluster ($r_c=4.4\arcsec$ in our
study). This may be explained by noticing that the cluster center here
determined is offset by almost $2\arcsec$ from the one derived by MG03
(which is also adopted by Mv05; see Fig. \ref{centers}). This is
indeed a large difference for such a compact cluster, and it is likely
the main reason for the detected discrepancy in the value of $r_c$. In
addition, we note that the King model fit to the surface brightness
profile of NGC 2210 is rather poor in Mv05 (see the $\chi^2$ values in
their Table 10). On the other hand, the two values of $r_c$ well agree
for NGC 2257, in spite of a $\sim 6\arcsec$ offset between the two
center estimates. This is because the cluster core is so large ($r_c
\sim 27\arcsec$) that an error in the determination of the center has
little impact on the resulting value of $r_c$.  With respect to S18,
the largest discrepancy is found for NGC 1466, while a reasonable
agreement is found for the other clusters. Also in this case, NGC 1466
is the most compact systems among the four in common with S18, and the
detected discrepancy may be due to a different location of the center,
that in S18 is offset by $\sim 6\arcsec$ toward the north, with
respect to our determination (Fig. \ref{centers}).  The comparison
with the effective and tidal radii quoted by Mv05 (Figure
\ref{cfr_rert}) shows only one notable discrepancy: for NGC 1898, the
Mv05 values for both the scale-lengths exceed those here determined by
factors of $\sim 2$.  As apparent from Figure \ref{cmd1898}, the LMC
field density contribution in the direction of this cluster is very
high. Hence, the detected discrepancy could be explained if Mv05
performed an insufficient subtraction of the background density from
the observed (surface brightness) profile. This would, in fact, induce
systematic overestimates of the large-scale radii (as $r_e$ and even
more $r_t$).

\begin{figure*}[!t]
\centering
\includegraphics[width=0.6\textwidth]{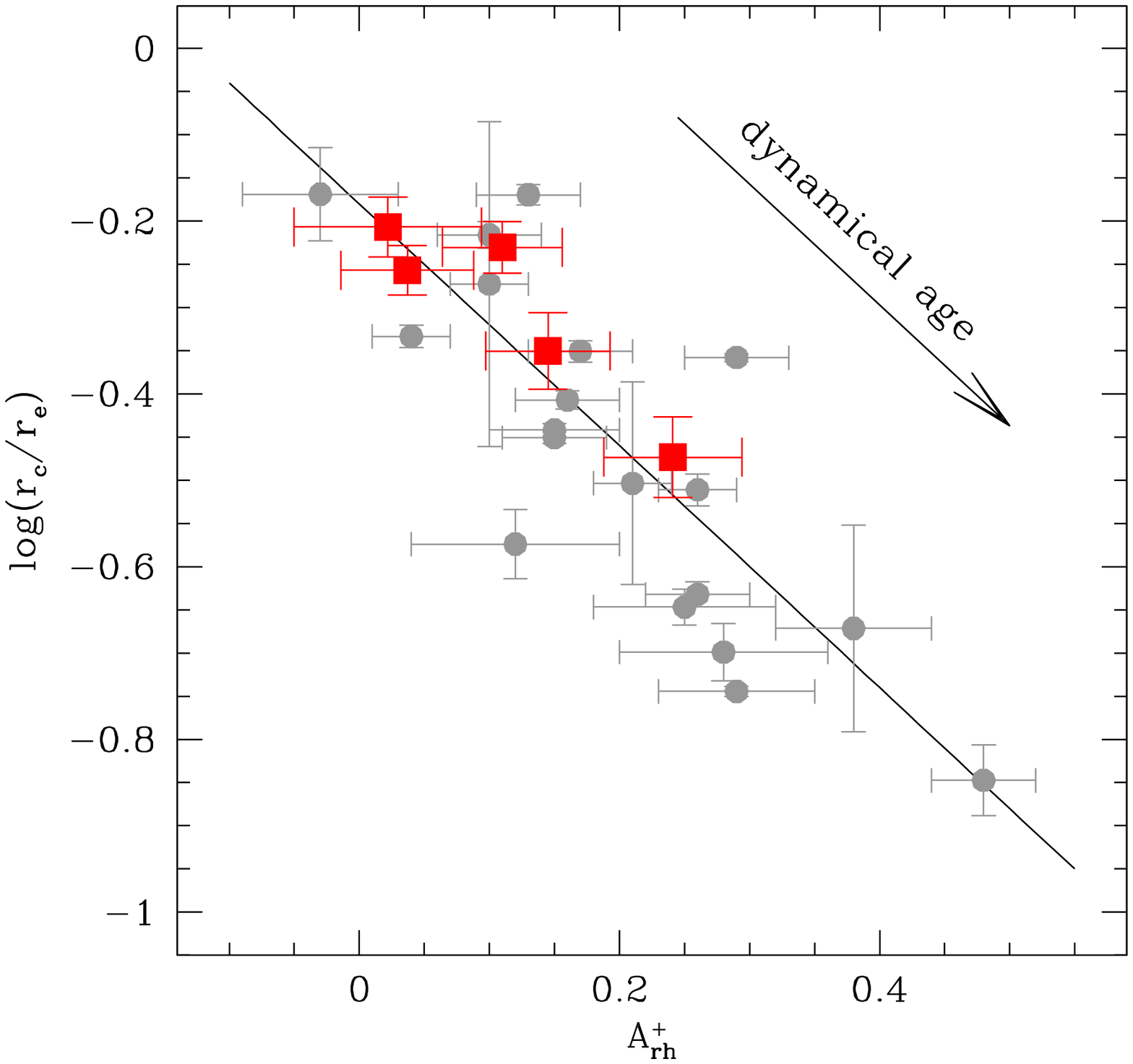}
\caption{Relation between the $A^+$ parameter and the ratio between
  the core and the effective radii for five of the six investigated
  LMC clusters (all, but NGC 1898; red squares), compared to that
  obtained for a sample of 19 globulars in the Milky Way (grey
  circles).  The $A^+$ parameter is a sensitive indicator of the level
  of internal dynamical evolution suffered by star clusters, its value
  increasing for increasing dynamical age of the system (see the
  arrow).}
\label{apiu}
\end{figure*}

As mentioned in the Introduction, \citet{ferraro+19} recently
investigated the dynamical ages of five of the surveyed clusters
(namely, NGC 1466, NGC 1841, NGC 2210, NGC 2257 and Hodge 11), by
measuring the central segregation of their BSS populations through the
$A^+$ parameter. The authors found that these five LMC old clusters
reached different levels of dynamical evolution and nicely follow the
relation between $A^+$ and $r_c$ drawn by the Galactic GC population.
The trend is consistent with the expectations of a long-term evolution
driven by two-body relaxation, with $r_c$ progressively decreasing in
time. In turn, this implies that the large spread in core radii
observed in coevally old star clusters in the LMC (e.g., MG03) can be
interpreted as the natural effect of their internal dynamical
evolution, which progressively moves stellar systems with relaxation
times significantly shorter than their ages toward small $r_c$
configurations.  In such a scenario, no extra energy sources
(provided, e.g., by a significant population of binary black holes;
\citealp{mackey+08}) are required to explain the observed age-$r_c$
distribution in the LMC, and there is no more need of an evolutionary
path where compact young clusters evolve into old globulars with a
wide range of core radii (see the discussion in \citealp{ferraro+19}).

The present study allows us to further investigate this issue by
studying the link between the dynamical indicator $A^+$ and
$r_c/r_e$. In fact, this ratio is expected to progressively decrease
with the long-term evolution of the cluster, unless an efficient
energy source intervenes to halt the core contraction and possibly
even induce core expansion \citep[e.g.][]{merritt+04, baumg05,
  mackey+08, trenti+10}. Figure \ref{apiu} shows the behavior of
$r_c/r_e$ as a function of the $A^+$ parameter for the 5 LMC clusters
in common with \citet[][red squares]{ferraro+19}, and for a sample of
19 Galactic GCs (grey circles) for which $A^+$ and $r_e$ have been
homogeneously measured (see \citealt{miocchi+13, lanzoni+16,
  ferraro+18b}). The observed trend is in perfect agreement with what
shown in Figure 8 of \citet{miocchi+13} for Milky Way
GCs,\footnote{Note that, instead of $A^+$, \citet{miocchi+13} adopted
  a different dynamical evolution indicator, namely, the position of
  the minimum of the normalized BSS radial distribution. This is, in
  fact, an alternative (but consistent) dynamical age indicator, as
  extensively discussed in \citet{lanzoni+16}.} and it indicates that
systems characterized by large values of $r_c/r_e$ are dynamically
younger than those showing small values of this ratio, as expected
from a two-body relaxation driven evolution.  Figure \ref{apiu} also
shows that no GCs with large dynamical age have a large value of
$r_c/r_e$, thus further supporting the conclusion that systems with
large core radii are just dynamically younger than the more compact
ones, with no need of invoking anomalous energy sources able to induce
core expansion. Of course, the fraction of massive dark remnants
(black holes and neutron stars) retained within the cluster potential
well has a significant impact in delaying the time-scale of the BSS
sedimentation process (e.g., \citealt{alessandrini16}), and the
measure of $A^+$ cannot discriminate between the presence or the
absence of a large population of these objects.  However, the
observational evidence of significant fractions of black holes in GCs
is still sparse. Therefore, it appears that the main driver of the BSS
segregation process (and of the resulting value of $A^+$) is the
long-term internal dynamical evolution of the system, while the action
of dark remnants can be considered at most as a second-order effect.
At this stage, the measure of the dynamical age of the most compact
old LMC clusters is of utmost importance to fully confirm this
scenario and to identify any system that might deviate from the trend
shown in Figure \ref{apiu}.

\acknowledgments Based on observations with the NASA/ESA
\textit{Hubble Space Telescope}, obtained at the Space Telescope
Science Institute, which is operated by AURA, Inc., under NASA
contract NAS 5-26555.  This paper is part of the project {\it
  Cosmic-Lab} (``Globular Clusters as Cosmic Laboratories'') at the
Physics and Astronomy Department of the Bologna University (see the
web page: \texttt{http//www.cosmic-lab.eu/Cosmic-Lab/Home.html}). The
research is funded by the project {\it Dark-on-Light} granted by MIUR
through PRIN2017 contract (PI: Ferraro).

\facilities{\textit{HST}(ACS/WFC, WFC3/UVIS)}
\software{\texttt{DAOPHOT} (\citealt{stetson94})}.


\begin{thebibliography}{}

\bibitem[Alessandrini et al.(2016)]{alessandrini16} Alessandrini, E.,
  Lanzoni, B., Miocchi, P., Ferraro, F.~R., \& Vesperini, E.\ 2016,
  \apj, 833, 252

\bibitem[Anderson et al.(2008)]{anderson08} Anderson, J., Sarajedini,
  A., Bedin, L.~R., et al.\ 2008, \aj, 135, 2055

\bibitem[Baumgardt et al.(2005)]{baumg05} Baumgardt, H., Makino, J.,
  \& Hut, P.\ 2005, \apj, 620, 238
  
\bibitem[Baumgardt, \& Hilker(2018)]{baum_hilker18} Baumgardt, H., \&
  Hilker, M.\ 2018, \mnras, 478, 1520

\bibitem[Beccari et al.(2010)]{beccari10} Beccari, G., Pasquato, M., De Marchi, G., et al.\ 2010, \apj, 713, 194

\bibitem[Beccari et al.(2019)]{beccari+19} Beccari G., Ferraro, F.R., Dalessandro, E., et al.\ 2019, \apj, 876, 87

\bibitem[Bellazzini et al.(2002)]{bellazzini02} Bellazzini, M., Fusi Pecci, F., Messineo, M., et al.\ 2002, \aj, 123, 1509

\bibitem[Calzetti et al.(1993)]{calzetti+93} Calzetti, D., de Marchi, G., Paresce, F., \& Shara, M.\ 1993, \apjl, 402, L1
  
\bibitem[Cadelano et al.(2017)]{cadelano+17} Cadelano, M.,
  Dalessandro, E., Ferraro, F.~R., et al.\ 2017, \apj, 836, 170

\bibitem[Dalessandro et al.(2008)]{ema08} Dalessandro, E., Lanzoni, B., Ferraro, F.~R., et al.\ 2008, \apj, 681, 311

\bibitem[Dalessandro et al.(2013)]{ema13} Dalessandro, E., Ferraro, F.~R., Massari, D., et al.\ 2013, \apj, 778, 135

\bibitem[Dalessandro et al.(2015)]{ema15} Dalessandro, E., Ferraro, F.~R., Massari, D., et al.\ 2015, \apj, 810, 40
  
\bibitem[Dalessandro et al.(2018)]{ema18} Dalessandro, E., Cadelano, M., Vesperini, E., et al.\ 2018, \apj, 859, 15 

\bibitem[Dubath et al.(1997)]{dubath+97} Dubath, P., Meylan, G., \& Mayor, M.\ 1997, \aap, 324, 505

\bibitem[Elson et al.(1989)]{EFF} Elson, R.~A.~W., Fall, S.~M., \& Freeman, K.~C.\ 1989, \apj, 336, 734

\bibitem[Ferraro et al.(1997)]{ferraro+97} Ferraro, F.~R., Paltrinieri, B., Fusi Pecci, F., et al.\ 1997, \aap, 324, 915 

\bibitem[Ferraro et al.(1999)]{ferraro+99} Ferraro, F.~R., Paltrinieri, B., Rood, R.~T., \& Dorman, B.\ 1999, \apj, 522, 983 

\bibitem[Ferraro et al.(2003)]{ferraro+03} Ferraro, F.~R., Possenti, A., Sabbi, E., et al.\ 2003, \apj, 595, 179 

\bibitem[Ferraro et al.(2006)]{ferraro+06} Ferraro, F.~R., Sabbi, E., Gratton, R., et al.\ 2006, \apjl, 647, L53

\bibitem[Ferraro et al.(2009)]{ferraro+09} Ferraro, F.~R., Beccari, G., Dalessandro, E., et al.\ 2009, \nat, 462, 1028 

\bibitem[Ferraro et al.(2012)]{ferraro+12} Ferraro, F.~R., Lanzoni, B., Dalessandro, E., et al.\ 2012, \nat, 492, 393 

\bibitem[Ferraro et al.(2018b)]{ferraro+18b} Ferraro, F.~R., Lanzoni,
  B, Raso, S., et al.\ 2018b, \apj, 860, 36

\bibitem[Ferraro et al.(2018a)]{ferraro+18a} Ferraro, F.~R.,
  Mucciarelli, A., Lanzoni, B., et al.\ 2018a, \apj, 860, 50

\bibitem[Ferraro et al.(2019)]{ferraro+19} Ferraro, F.~R., Lanzoni, B., Dalessandro, E. et al.\ 2019, Nature Astronomy, in press

\bibitem[Fiorentino et al.(2014)]{fiorentino14} Fiorentino, G., Lanzoni, B., Dalessandro, E., et al.\ 2014, \apj, 783, 29

\bibitem[Harris(1996)]{h96} Harris, W.~E.\ 1996, \aj, 112, 1487, 2010 edition

\bibitem[King(1966)]{king66} King, I.~R.\ 1966, \aj, 71, 64 

\bibitem[Lanzoni et al.(2007a)]{lanzoni+07a} Lanzoni, B., Dalessandro, E., Ferraro, F.~R., et al.\ 2007a, \apj, 663, 267

\bibitem[Lanzoni et al.(2007b)]{lanzoni+07b} Lanzoni, B., Sanna, N., Ferraro, F.~R., et al.\ 2007b, \apj, 663, 1040

\bibitem[Lanzoni et al.(2007c)]{lanzoni+07c} Lanzoni, B., Dalessandro, E., Ferraro, F.~R., et al.\ 2007c, \apjl, 668, L139 

\bibitem[Lanzoni et al.(2010)]{lanzoni+10} Lanzoni, B., Ferraro, F.~R., Dalessandro, E., et al.\ 2010, \apj, 717, 653 

\bibitem[Lanzoni et al.(2013)]{lanzoni+13} Lanzoni, B., Mucciarelli, A., Origlia, L., et al.\ 2013, \apj, 769, 107

\bibitem[Lanzoni et al.(2016)]{lanzoni+16} Lanzoni, B., Ferraro, F.~R., Alessandrini, E., et al.\ 2016, \apjl, 833, L29

\bibitem[Lanzoni et al.(2018a)]{lanzoni+18a} Lanzoni, B., Ferraro, F.~R., Mucciarelli, A., et al.\ 2018a, \apj, 865, 11 

\bibitem[Lanzoni et al.(2018b)]{lanzoni+18b} Lanzoni, B., Ferraro, F.~R., Mucciarelli, A., et al.\ 2018b, \apj, 861, 16 

\bibitem[Lugger et al.(1995)]{lugger+95} Lugger, P.~M., Cohn, H.~N., \& Grindlay, J.~E.\ 1995, \apj, 439, 191 

\bibitem[L{\"u}tzgendorf et al.(2011)]{lutzgendorf+11} L{\"u}tzgendorf, N., Kissler-Patig, M., Noyola, E., et al.\ 2011,
  \aap, 533, A36

\bibitem[Mackey \& Gilmore (2003)]{mackey03} Mackey, A.D., Gilmore,G.F.,\ 2003, \mnras, 338, 85 (MG03)

\bibitem[Mackey et al. (2008)]{mackey+08} Mackey, A.D., Wilkinson, M. I., Davies, M. B., et al.\ 2008, \mnras, 386, 65

\bibitem[McLaughlin \& van der Marel(2005)]{mclaugh05} McLaughlin, D.~E., \& van der Marel, R.~P.\ 2005, \apjs, 161, 304 (Mv05)

\bibitem[Merritt et al.(2004)]{merritt+04} Merritt, D., Piatek, S.,
  Portegies Zwart, S., et al.\ 2004, \apjl, 608, L25

\bibitem[Meylan, \& Heggie(1997)]{meylan+97} Meylan, G., \& Heggie, D.~C.\ 1997, \aapr, 8, 1

\bibitem[Miocchi et al.(2013)]{miocchi+13} Miocchi, P., Lanzoni, B., Ferraro, F.~R., et al.\ 2013, \apj, 774, 151 

\bibitem[Montegriffo et al.(1995)]{montegriffo+95} Montegriffo, P., Ferraro, F.~R., Fusi Pecci, F., \& Origlia, L.\ 1995, \mnras, 276, 739 

\bibitem[Noyola, \& Gebhardt(2006)]{noyola+06} Noyola, E., \& Gebhardt, K.\ 2006, \aj, 132, 447
  
\bibitem[Piatti et al.(2019)]{piatti19} Piatti, A.~E., Webb, J.~J., \&
  Carlberg, R.~G.\ 2019, \mnras, 489, 4367

\bibitem[Pietrzy{\'n}ski et al.(2013)]{pietrzynski13} Pietrzy{\'n}ski, G., Graczyk, D., Gieren, W., et al.\ 2013, \nat, 495, 76 

\bibitem[Raso et al.(2017)]{raso+17} Raso, S., Ferraro, F.~R., Dalessandro, E., et al.\ 2017, \apj, 839, 64

\bibitem[Raso et al.(2019)]{raso+19} Raso, S., Pallanca, C., Ferraro, F.~R., et al.\ 2019, \apj, 879, 56
  
\bibitem[Salinas et al.(2012)]{salinas+12} Salinas, R., J{\'\i}lkov{\'a}, L., Carraro, G., et al.\ 2012, \mnras, 421, 960

\bibitem[Saracino et al.(2015)]{saracino+15} Saracino, S., Dalessandro, E., Ferraro, F.~R., et al.\ 2015, \apj, 806, 152 

\bibitem[Shara et al.(1997)]{shara97} Shara, M.~M., Saffer, R.~A., \& Livio, M.\ 1997, \apjl, 489, L59

\bibitem[Sollima et al.(2017)]{sollima+17} Sollima, A., Dalessandro,
  E., Beccari, G., \& Pallanca, C.\ 2017, \mnras, 464, 3871
  
\bibitem[Stetson(1994)]{stetson94} Stetson, P.~B. 1994, \pasp, 106, 250

\bibitem[Sun et al.(2018)]{sun18} Sun, W., Li, C., de Grijs, R., et al.\ 2018, \apj, 862, 133 (S18)

\bibitem[Trenti et al.(2010)]{trenti+10} Trenti, M., Vesperini, E., \& Pasquato, M.\ 2010, \apj, 708, 1598

\end{thebibliography}
\end{document}